\documentclass[aps,prm,,showpacs,showkeys,preprint,superscriptaddress,longbibliography,floatfix]{revtex4-1}
\usepackage{graphicx}
\usepackage{dcolumn}
\usepackage{bm}
\usepackage{mhchem}
\usepackage{color}
\usepackage{textgreek}
\usepackage[T1]{fontenc}
\usepackage[utf8]{inputenc}

\usepackage[normalem]{ulem}

\begin{document}

\title{Growth and Structure of Singly-Oriented Single-Layer Tungsten Disulfide on Au(111)}
\author{Luca Bignardi}
\author{Daniel Lizzit}
\affiliation{Elettra - Sincrotrone Trieste S.C.p.A., AREA Science Park, Strada Statale 14, km 163.5, 34149 Trieste, Italy.}

\author{Harsh Bana}
\author{Elisabetta Travaglia}
\affiliation{Department of Physics, University of Trieste, Via Valerio 2, 34127 Trieste, Italy.}

\author{Paolo Lacovig}
\affiliation{Elettra - Sincrotrone Trieste S.C.p.A., AREA Science Park, Strada Statale 14, km 163.5, 34149 Trieste, Italy.}

\author{Charlotte E. Sanders}
\author{Maciej Dendzik}
\altaffiliation{Current address: Department of Physical Chemistry, Fritz-Haber-Institut of the Max Planck Society, Faradayweg 4-6, Berlin 14915, Germany. }
\author{Matteo Michiardi}
\author{Marco Bianchi}
\affiliation{Department of Physics and Astronomy, Interdisciplinary Nanoscience Center (iNANO), Aarhus University, Ny Munkegade 120, 8000 Aarhus C, Denmark.}

\author{Moritz Ewert}
\affiliation{Institute of Solid State Physics, University of Bremen, Otto-Hahn-Allee 1, 28359 Bremen, Germany}
\affiliation{MAPEX Center for Materials and Processes, Bremen, Germany}
\author{Lars Bu\ss}
\affiliation{Institute of Solid State Physics, University of Bremen, Otto-Hahn-Allee 1, 28359 Bremen, Germany}
\author{Jens Falta}
\author{Jan Ingo Flege}
\affiliation{Institute of Solid State Physics, University of Bremen, Otto-Hahn-Allee 1, 28359 Bremen, Germany}
\affiliation{MAPEX Center for Materials and Processes, Bremen, Germany}

\author{Alessandro Baraldi}
\affiliation{Elettra - Sincrotrone Trieste S.C.p.A., AREA Science Park, Strada Statale 14, km 163.5, 34149 Trieste, Italy.}
\affiliation{Department of Physics, University of Trieste, Via Valerio 2, 34127 Trieste, Italy.}
\affiliation{IOM-CNR, Laboratorio TASC, AREA Science Park, Strada Statale 14, km 163.5, 34149 Trieste, Italy.}

\author{Rosanna Larciprete}
\affiliation{CNR-Institute for Complex Systems, Via dei Taurini 19, 00185 Roma, Italy.}

\author{Philip Hofmann}
\email{philip@phys.au.dk}
\affiliation{Department of Physics and Astronomy, Interdisciplinary Nanoscience Center (iNANO), Aarhus University, Ny Munkegade 120, 8000 Aarhus C, Denmark.}

\author{Silvano Lizzit}
\email{lizzit@elettra.eu}
\affiliation{Elettra - Sincrotrone Trieste S.C.p.A., AREA Science Park, Strada Statale 14, km 163.5, 34149 Trieste, Italy.}

\begin{abstract}
We present a complete characterization at the nanoscale of the growth and  structure of single-layer tungsten disulfide (\ce{WS2}) epitaxially grown on Au(111). Following the growth process in real time with fast x-ray photoelectron spectroscopy, we obtain a singly-oriented layer by choosing the proper W evaporation rate and substrate temperature during the growth. Information about the morphology, size and layer stacking of the \ce{WS2} layer were achieved by employing x-ray photoelectron diffraction and low-energy electron microscopy. The strong spin splitting in the valence band of \ce{WS2} coupled with the single-orientation character of the layer make this material the ideal candidate for the exploitation of the spin and valley degrees of freedom.  
\end{abstract}


\maketitle

\section{Introduction}
Only a short time after the successful creation of graphene by mechanical exfoliation \cite{Novoselov:2004aa,Novoselov:2005aa,Zhang:2005ab}, it was realized that other two-dimensional materials could be obtained in a similar way from layered crystals, especially  transition metal dichalcogenides (TMDCs) \cite{Novoselov:2005ab}. This has lead to an intense study of single layer (SL) TMDCs which, like graphene, show electronic properties that are remarkably different from those of their parent bulk compound. Initially, much attention had been given to the presence of a direct band gap  in some of the semiconducting SL TMDCs \cite{Splendiani:2010aa,Mak:2010aa} and the possibility to exploit this property in electronic and opto-electronic applications \cite{Radisavljevic:2011aa,Lopez-Sanchez:2013aa}. Later, it was suggested that these materials could be used for entirely new concepts in electronics because of their coupled spin \cite{Xiao:2012ab} and valley \cite{Mak:2012aa,Zeng:2012aa,Cao:2012ab,Mak:2014aa} degrees of freedom  \cite{Xu:2014ac}, something that does not play a role in graphene where inversion symmetry prevents the lifting of the spin degeneracy. 

The technological exploitation of the SL TMDCs will benefit from  production by bottom-up growth, rather than exfoliation. Chemical vapor deposition and molecular beam epitaxy appear to be promising approaches to obtain a high degree of orientation and a closed layer \cite{Lee:2012ac,Shi:2012aa,Li:2015ac,Pyeon:2016ih,Lehtinen:2015aa,Chen:2016dd,Fu:2017aa,Kang:2018dq}
An additional requirement for devices that shall make use of the valley degree of freedom, such as a valley filter \cite{Rycerz:2007aa}, is to avoid the presence of mirror domains, a common defect in the growth of SL TMDCs\cite{Lehtinen:2015aa}. The two mirrored versions of the SL TMDC unit cell have the same reciprocal lattice, with merely the K and -K points of the surface Brillouin zone inverted \cite{Bana2018} and the same band structure, apart from an inverted spin polarization for the two mirror domains. The presence of mirrored domains is thus detrimental for both fundamental studies on the coupled spin and valley degrees of freedom and for the use of TMDCs in devices \cite{Mak:2014aa,Mak2016}. 

There has been recent progress in the growth of single orientation two-dimensional materials, for example for SL BN on Ir(111) \cite{Orlando2014}, as well as for SL MoS$_2$ on hexagonal boron nitride \cite{Fu:2017aa} and Au(111) \cite{Bana2018}. Herein, we investigate the growth and structural features of epitaxially-grown SL \ce{WS2} on Au(111). We reveal that a bottom-up approach can be employed to obtain SL \ce{WS2} with a single orientation, a feature that has not been reported so far. This aspect, together with the spin-splitting in the valence band, which is one of the largest among TMDCs \cite{Zhu2011,Wang2012,Kormanyos:2015aa}, opens up the possibility to investigate and exploit the coupled valley and spin degrees of freedom in devices applications.  We characterize the samples by X-ray photoelectron spectroscopy (XPS) and determine the structural parameters and the extent and orientation of the crystalline domains by combining low-energy electron diffraction (LEED) and microscopy (LEEM) with X-ray photoelectron diffraction (XPD). XPD measurements were able to conclusively prove that SL \ce{WS2} can grow with a single orientation and determine the relative orientation of SL \ce{WS2} with respect to the Au(111) substrate. 

\section{Experimental Methods}
Growth and characterization experiments were carried out at the SuperESCA beamline\cite{Baraldi:2003ab} of the Elettra Synchrotron radiation facility in Trieste, Italy, except for the LEEM data, which were acquired at the University of Bremen.  The Au(111) substrate was cleaned by cycles of \ce{Ar+} ion sputtering at 2 keV and annealing up to 950~K. Cleanliness of the surface was checked by LEED and XPS, which did not detect traces of contamination. 
High-resolution S~$2p$ and W~$4f$ core level spectra were measured at room temperature in normal emission on the as-grown WS$_2$ single layer, using photon energies of 260 eV and 140 eV, respectively. The overall energy resolution was below 50~meV. Thanks to the combination of a high-flux beamline and a high-efficiency electron detector, it was possible to measure fast X-ray photoelectron spectroscopy (XPS) spectra during the growth, greatly aiding the optimization of growth parameters. 

Each XPD pattern was measured by collecting XPS spectra for more than 1000 different polar ($\theta$) and azimuthal ($\phi$) angles, as defined in the diagram in Figure \ref{fig:2}.
For each of these spectra, a peak fit analysis was performed (parameters of the analysis reported in Supplementary Material) and the intensity $I(\theta$, $\phi)$ of each component resulting from the fit, \textit{i.e.} the area under the photoemission line, was extracted. The resulting XPD patterns are the stereographic projection of the modulation function $\chi$, which was obtained from the peak intensity for each emission angle  $(\theta, \phi) $ as 
\begin{equation}
\chi=\frac{I(\theta, \phi) - I_0(\theta)}{I_0(\theta)},
\label{eq:chi}\end{equation}
 where $I_0$($\theta$) is the average intensity for each azimuthal scan at polar angle $\theta$. The structural determination was performed by comparing measured XPD patterns to multiple scattering simulations for a trial structure. Such patterns were simulated using the program package for Electron Diffraction in Atomic Clusters (EDAC)\cite{Garcia_2001}. The presence of two mirror domain orientations was taken into account by an incoherent superposition of the calculated intensities, based on the well-funded \cite{Gronborg:2015aa,Dendzik:2015aa} assumption of domain sizes that are sufficiently large to neglect boundary effects.  The agreement between the simulations and the experimental results was quantified by computing the reliability factor $(R)$ \cite{Woodruff:2007aa}
\begin{equation}R = \frac{\sum_{i} (\chi_{\text{exp}, i} - \chi_{\text{sim}, i})^2}{\sum_{i} ({\chi^2}_{\text{exp}, i} + {\chi^2}_{\text{sim}, i})},\label{eq:R}\end{equation} where  $\chi_{\text{sim}, i}$ and  $\chi_{\text{exp}, i}$ are the simulated and the experimental modulation functions for each emission angle $i$.
The estimation of the accuracy of the quantities derived by means of R-factor analysis, i.e. lattice constant, layer thickness and percentage of mirror domains, was deduced from the R-factor confidence interval defined as  \cite{Pendry_1980}
\begin{equation}
\Delta R_{\text{min}}=R_{\text{min}}\sqrt{\frac{2}{N}},
\label{eq:ConfidenceInterval}
\end{equation}
where $R_{\text{min}}$ is the minimum R-factor value and $N$ is the number of well-resolved peaks in the XPD pattern $(N\sim350)$.
The LEED experiments were carried out using a commercial VG instrument installed at the experimental chamber of the SuperESCA beamline. 

LEEM data were recorded using an ELMITEC LEEM III instrument  at the University of Bremen. A sample was prepared at the SuperESCA beamline,  transferred to Bremen through air and then annealed to 670~K after inserting it in the ultra-high vacuum (UHV) chamber. 
As a diffraction based method, LEEM can access either real space (imaging mode) or reciprocal space (diffraction mode) providing information on both surface morphology and atomic structure.\cite{Flege2012}
By constricting the illuminated area with an aperture in diffraction mode, information on the local atomic structure can be obtained from areas as small as 250\,nm in diameter, an approach known as $\mu$LEED. 
In BF imaging contrast usually arises from differences in local atomic structure and composition. 
In DF-LEEM diffraction contrast is obtained since only regions contributing to this very beam will show up bright in the image. In LEEM, the \mbox{$I(V)$ curves} were extracted from a stack of images, yielding individual \mbox{$I(V)$ curves} for every image pixel, thus enabling structural identification on the nanoscale.\cite{Flege2014,Flege2015}. The LEED calculations were performed using the AQuaLEED package, which is based on the Barbieri/Van Hove SATLEED package and on the layer stacking implementation by N. Materer \cite{aquaLEED,barbieri,Materer2001}. The $I(V)$ curves have been calculated considering SL \ce{WS2} with the structural parameters obtained from XPD stacked on the Au(111) substrate and stretching the Au(111) surface lattice constant to match that one of \ce{WS2}.

 \section{Results and Discussion}
 \begin{figure}[!t]
 \includegraphics[width=\linewidth]{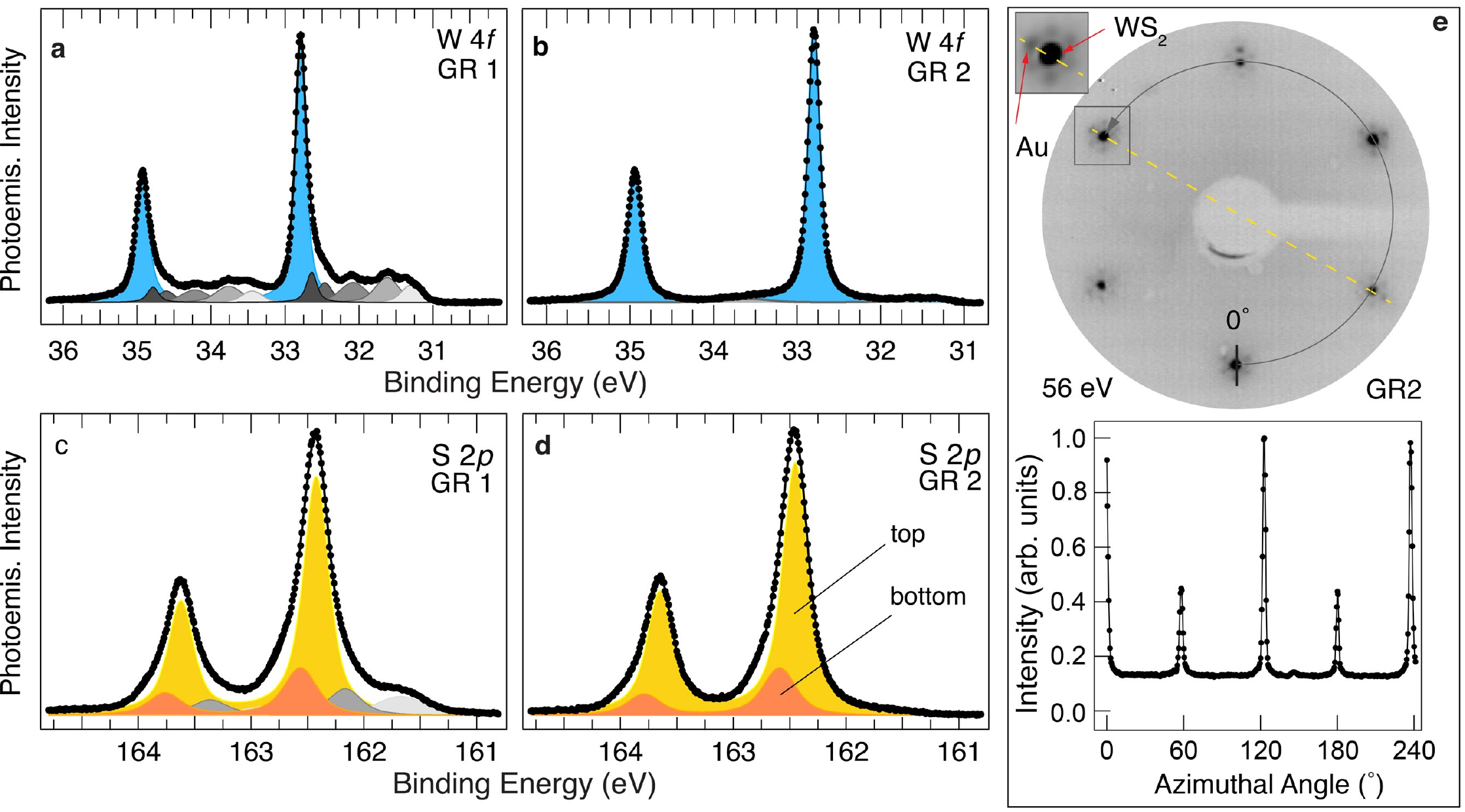}
\caption{(a), (b) High-resolution core level spectra for W $4f$ measured at room temperature (RT) after  GR1 and  GR2 (photon energy $h\nu=140$~eV). The result of a peak fitting is displayed as well. (c), (d) High-resolution core level spectra for S~$2p$ measured at RT after  GR1 and  GR2, respectively ($h\nu=260$~eV). (e) LEED pattern acquired on GR2 (primary electron energy=56~eV. The dashed yellow line indicates the alignment of the moiré superstructure with the crystallographic orientation of \ce{WS2}. The solid grey line is a circular profile taken across the \ce{WS2} spots and shown at the bottom of the panel.  }
  \label{fig:1}

\end{figure}
We have characterized SL \ce{WS2} on Au(111) prepared according to two distinct growth procedures, hereinafter labeled as GR1, that led to the growth of a mixture of two mirror domains, and GR2, which resulted in the growth of SL \ce{WS2} with a single orientation. In both growth methods, tungsten was dosed on Au(111) from a hot W filament in a background pressure of H$_2$S. For GR1, the substrate temperature was 873~K and the W deposition rate, measured by means of a quartz microbalance, was $7.5 \times 10^{-3}$~ML/min, with 1~ML corresponding to the surface atomic density of the Au(111) surface. The H$_2$S pressure was $1\times10^{-6}$~mbar. The growth conditions for GR1 are comparable to those commonly employed in the growth of \ce{WS2} on noble metal substrates \cite{Dendzik_2015}.
For  GR2 the growth was performed at a higher  sample temperature (923~K) while a more than two-times lower W deposition rate ($3.3\times10^{-3}$ ML/minute) was employed. \ce{H2S} pressure was set to  $1\times10^{-5}$~mbar. The growth was followed in real-time by fast-XPS, acquiring the W~$4f$ core level during the entire process. A full account of these measurements is presented in Supplementary Material.

 High-resolution W $4f$ core-level spectra  measured after  GR1 and GR2 are shown in Figure\,\ref{fig:1}a and b, respectively.  For GR1, the main spin-orbit doublet, interpreted as stemming from SL WS$_2$ (blue component)\cite{Dendzik_2015}, is accompanied by many small contributions (grey components) at lower binding energy (BE), attributed to incompletely sulfided W species of the form \ce{WS_{2-x}} $(0<x<1)$ or metallic W clusters, as observed earlier for the same system \cite{Dendzik_2015,Shpak2010,Fuchtbauer2013}. For GR2, on the other hand, these additional components are  entirely absent and the  W~4$f_{7/2}$ spectrum shows a single peak at $\text{BE}=32.78$~eV, characteristic of SL \ce{WS2}\cite{Dendzik_2015}, with its spin-orbit doublet separated by 2.14~eV.  The line-shape parameters used for the spectral deconvolution are given in Supplementary Material.

In Figure~\ref{fig:1}c and \ref{fig:1}d we show the S $2p$ spectra resulting from the two growth procedures. 
Two main components are visible in each of the spin-orbit peaks, indicated in light (S~$2p_{3/2}$ BE=162.45~eV) and dark (S~$2p_{3/2}$ BE=162.59~eV) orange: the latter has been assigned to S atoms from the `bottom'  (close to the Au(111) surface) while the former stems from the `top' atoms in the S-W-S sandwich arrangement of \ce{WS2}. This interpretation follows the assignments for SL \ce{MoS2} prepared on Au(111) \cite{Bana2018}, and it will be discussed in the XPD experiments aiming to determine the polytype of the SL (see Supplementary Material). 
Besides these two components we observed several extra peaks at lower binding energy for GR1 , which are originating from species other than \ce{WS2}.  On the contrary, the spectrum acquired on GR2 shows the spectral components associated with top and bottom sulfur only, indicating a better quality of the grown layer.

Figure\,\ref{fig:1}e reports the LEED pattern acquired on the GR2  sample. 
The most intense spots corresponding to the reciprocal lattice of \ce{WS2} are surrounded by moir\'{e} satellites, due to the lattice mismatch between \ce{WS2} and the Au substrate\cite{Dendzik_2015}. 
The first order Au(111) diffraction spots also exhibit a hexagonal pattern, though slightly bigger in dimension due to a smaller lattice vector, and they are aligned with those of \ce{WS2} (see dashed yellow line in Figure~\ref{fig:1}e and inset thereinto). 
By comparing the reciprocal lattice vectors of \ce{WS2} with those of Au(111), we obtain a moir\'{e} periodicity of $3.19\pm0.1$~nm, in close agreement with literature \cite{Dendzik_2015}. This is indicative of the formation of a superstructure with a periodicity of $(10\times10)$-\ce{WS2} on $(11\times11)$-Au unit cell configuration. 

Furthermore, the LEED pattern acquired on GR2 displays a three-fold symmetry  (as evidenced by the circular profile taken across the \ce{WS2} spots in the pattern), which could indicate the presence, for this sample, of a single azimuthal orientation of the \ce{WS2} crystalline domains. However, a conclusive answer about the orientation of the layers and about the absence of detrimental mirror domains can only be obtained by means of XPD experiments.

\begin{figure*}[p]
\includegraphics[width=.85\linewidth]{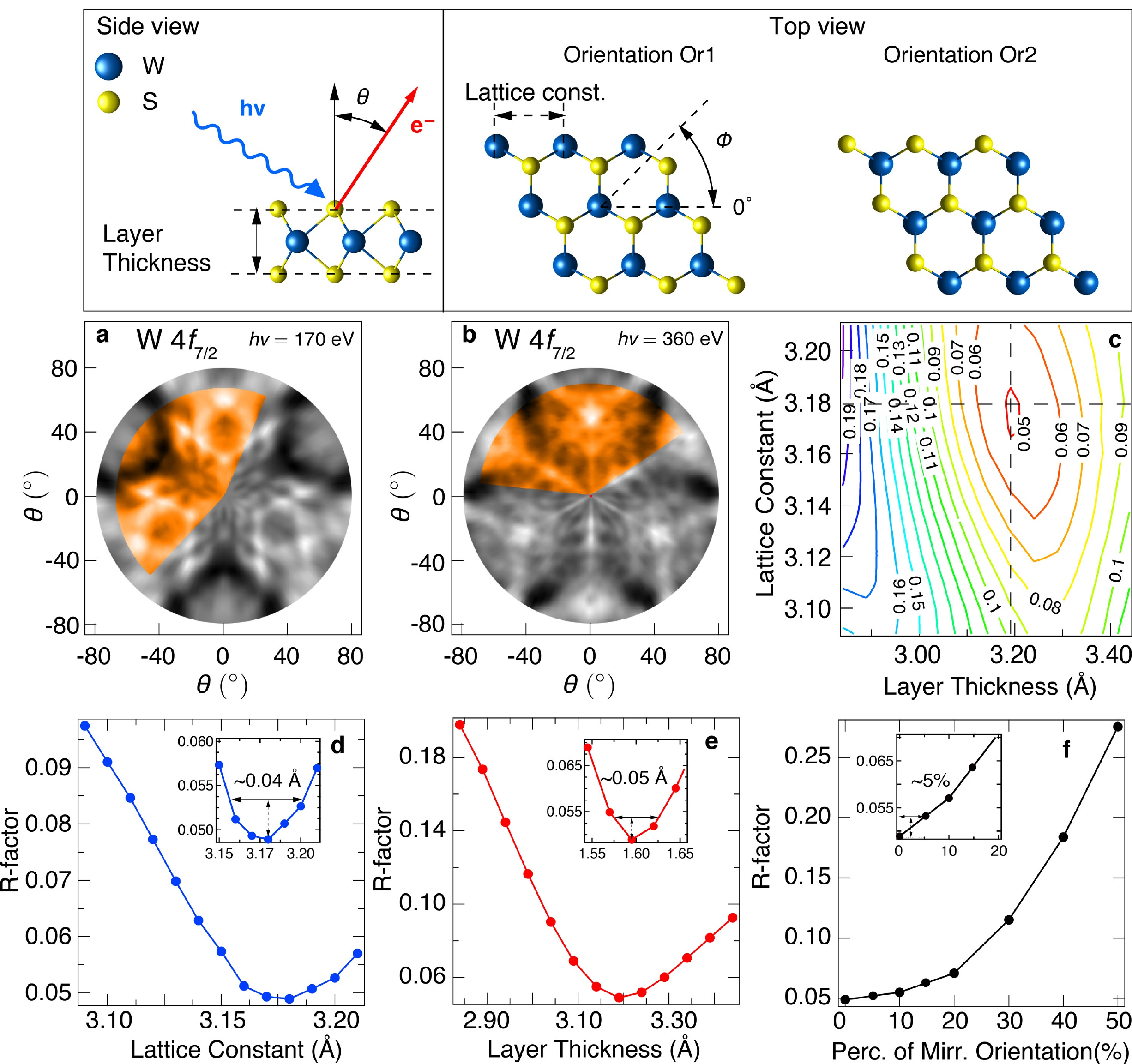}\\
\caption{(top-panel) Side view (left) of the trigonal prismatic (1H) structural phase of \ce{WS2}, and top view (right) of the main (Or1) and mirror (Or2) orientations used for XPD simulations.
 (a) XPD pattern acquired on sample GR2 for the W~$4f_{7/2}$ component ($h\nu=170$~eV) (in color) together  with a simulation assuming a single orientation of the layer (in greyscale). (b) W~$4f_{7/2}$ XPD pattern obtained on sample GR2, with $h\nu=360$~eV. (c) Contour plot reporting the R-factor as a function of the layer thickness and lattice constant. (d) and (e) R-factor along the dashed lines in panel (c).  (f) R-factor vs. percentage of mirror orientation admixture, obtained for layer thickness and lattice constant values corresponding to the minima of the R-factor shown in panels (d) and (e). 
The inset of (d), (e) and (f) panels shows a magnification of the graph around the minimum of the R-factor, with the vertical dashed arrow indicating the confidence interval $\Delta$R$_{\text{min}}$ and the horizontal arrow the uncertainty in the determination of the lattice constant, layer thickness and percentage of mirror orientation, respectively, as explained in Methods. The simulated pattern shown are those obtained employing the parameters deduced from the R-factor analysis. }
  \label{fig:2}
 
\end{figure*}
In XPD, photoemission intensity modulations arise from the interference between the component of the photoelectron wave field that reaches the detector directly from the emitting atom and the components scattered by atoms surrounding the emitter. 
This makes XPD very sensitive to the local environment of the emitter \cite{Woodruff:2007aa}, hence making this technique ideal to address questions about the polytype of the layer (trigonal prismatic or octahedral), the distribution of mirror domains and other structural parameters such as lattice constant and layer thickness. 

We acquired XPD patterns from the W~$4f_{7/2}$ core level on the layer produced with GR2 using a photon energy $h\nu=170$~eV (photoelectron kinetic energy, KE$\sim$137~eV), at which both forward and backward scattering contributions are significant, and at $h\nu=360$~eV (KE$\sim$326~eV), enhancing the forward scattering and suppressing the  backscattering\cite{Woodruff:2007aa}.  
The experimental XPD patterns (color sector in Figure\,\ref{fig:2}a and b) were compared to simulations (greyscale) using the reliability factor (R-factor) analysis (see Methods). Simulated diffraction patterns were calculated for different structural phases and their corresponding structural parameters assuming, as initial guess, the trigonal prismatic (1H) phase (sketched in Figure~\ref{fig:2}), following the indications obtained earlier for this system \cite{Dendzik_2015}. The presence of two orientations in the SL can be modeled by assuming an incoherent superposition from the domains, such that the total photoemission intensity is

\begin{equation}
I_{\text{tot}}=aI_{0}+bI_{\text{mir}}\qquad(b=1-a),
\label{eq:mirrdomains}
\end{equation}
where $I_0$ is the contribution to the XPD pattern from the main orientation (Or1) and $I_{\text{mir}}$ is the contribution from the mirror orientation (Or2), according to the models shown in top-right of Figure \ref{fig:2}. The simulated XPD patterns for the two mirrored orientations stemming from W~$4f_{7/2}$ (KE$\sim$326~eV) are reported in the Supplementary Material.

The three-fold symmetry of the simulated patterns for a single orientation matches the experimental data, which is a direct experimental indication of a dominant domain orientation and hints at a minor contribution from the mirror orientation. However, a conclusive determination of the layer orientation was achieved only through the R-factor analysis, as shown in the following. 
For the simulations of the photoemission intensity by multiple scattering calculations, the values for layer thickness (i.e. the vertical S-S distance), the lattice constant and the fraction of mirror domains in the SL, assuming the 1H phase, were changed independently, running an R-factor optimization in the three-dimensional parameter space, using the data sets of both Figure\,\ref{fig:2}a and b. 
The analysis clearly reveals a global minimum for a single orientation structure (a maximum of 5\% contribution of mirror domains), a lattice parameter of $3.17\pm0.04$~\AA~and a layer thickness of $3.17\pm0.05$~\AA~in very good agreement with the values reported in literature \cite{Schutte_1987}. The very low absolute value of the minimum R-factor ($R=0.05$) indicates an excellent agreement between experiment and simulation.  For the sake of simplicity, Figure\,\ref{fig:2}c reports a two-dimensional plot of the R-factor as a function of lattice constant and layer thickness obtained for the single orientation ($b=0$).  Figures\,\ref{fig:2}d and \ref{fig:2}e are  obtained by taking cuts along the dashed lines in panel (c) to highlight the R-factor trend around the absolute minimum, while Figure\,\ref{fig:2}f clearly shows that a minimum of R-factor is observed when only a single orientation is present. This is a considerably different from earlier reports on the growth of SL \ce{WS2} by bottom-up approaches \cite{Fuchtbauer2013}, in which the presence of detrimental mirror domains could not be avoided. 
 
\begin{figure}[t]
\begin{center}
\includegraphics[width=.6\linewidth]{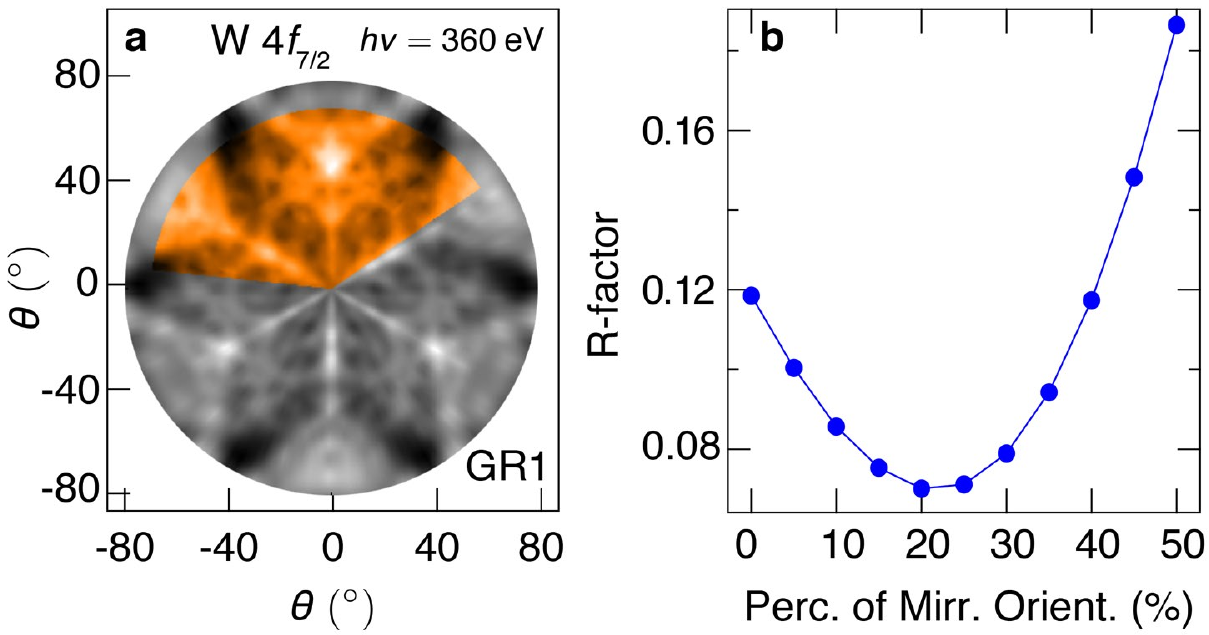}
\caption{(a) XPD experimental pattern (colored) and simulated pattern (greyscale) for the  W 4f${_{7/2}}$ photoemission intensity modulation  from sample GR1. The XPD pattern was measured with a photon energy of 360~eV, corresponding to a kinetic energy of 326~eV. (b) R-factor vs percentage of mirror orientation admixture for sample GR1.}
  \label{fig:3}
  \end{center}
\end{figure} 
A similar analysis on the XPD patterns from  S~$2p_{3/2}$ and  W~$4f_{7/2}$ was performed assuming a octahedral (1T) polytype for the \ce{WS2} SL. The simulations of the XPD patterns were performed using the same lattice constant and layer thickness determined earlier for the 1H polytype. The  analysis returned higher values of the R-factor for the 1T phase, ruling then out this structure. The details of this analysis are reported in the Supplementary Material. 

To understand whether also GR1 leads to a singly-oriented SL \ce{WS2}, we carried out an R-factor analysis simulating XPD patterns with a different percentage of mirror domains and compared them with the experimental pattern obtained from the W~$4f_{7/2}$ peak belonging to \ce{WS2} acquired on GR1 (Figure \ref{fig:3}a). The other structural parameters, obtained from the analysis on GR2, remained unchanged. In this case, the absolute minimum of the R-factor (0.07) was found when including  $\sim$20\% of mirror domains in the simulation, as shown in Figure \ref{fig:3}b. 

It is interesting to ask what physical mechanisms are responsible for the growth of the single orientation. 
The only element breaking the symmetry is the Au(111) substrate. While the first Au layer has a six-fold symmetry, considering deeper layers lowers the symmetry to three-fold, providing a suitable template for the growth of singly-oriented \ce{WS2}. However, this feature can only be exploited by an adequate choice of the growth parameters, in particular the W deposition rate, the substrate temperature and the \ce{H2S} partial pressure. Optimizing these does not only improve the quality of the resulting SL \ce{WS2} but drives the formation of a single domain orientation. This could be favorable to achieve in GR2 because a higher temperature allows an enhanced surface diffusion of W atoms on the Au substrate while a lower W deposition rate, together with a higher \ce{H2S} background pressure, avoids the formation of partially sulfided species, as seen by XPS.

Having established that  GR2 results in the growth of a single domain orientation, we used the XPD pattern from the W~$4f_{7/2}$ core-level component for GR2 to determine the orientation of the \ce{WS2} layer with respect to the Au(111) substrate. 
In order to do this, we determined the orientation of the Au(111) substrate by measuring an XPD pattern stemming from the bulk component of Au~$4f_{7/2}$ on the clean Au sample, i.e. before the deposition of \ce{WS2}.
\begin{figure*}[p]
\begin{center}
\includegraphics[width=0.65\linewidth]{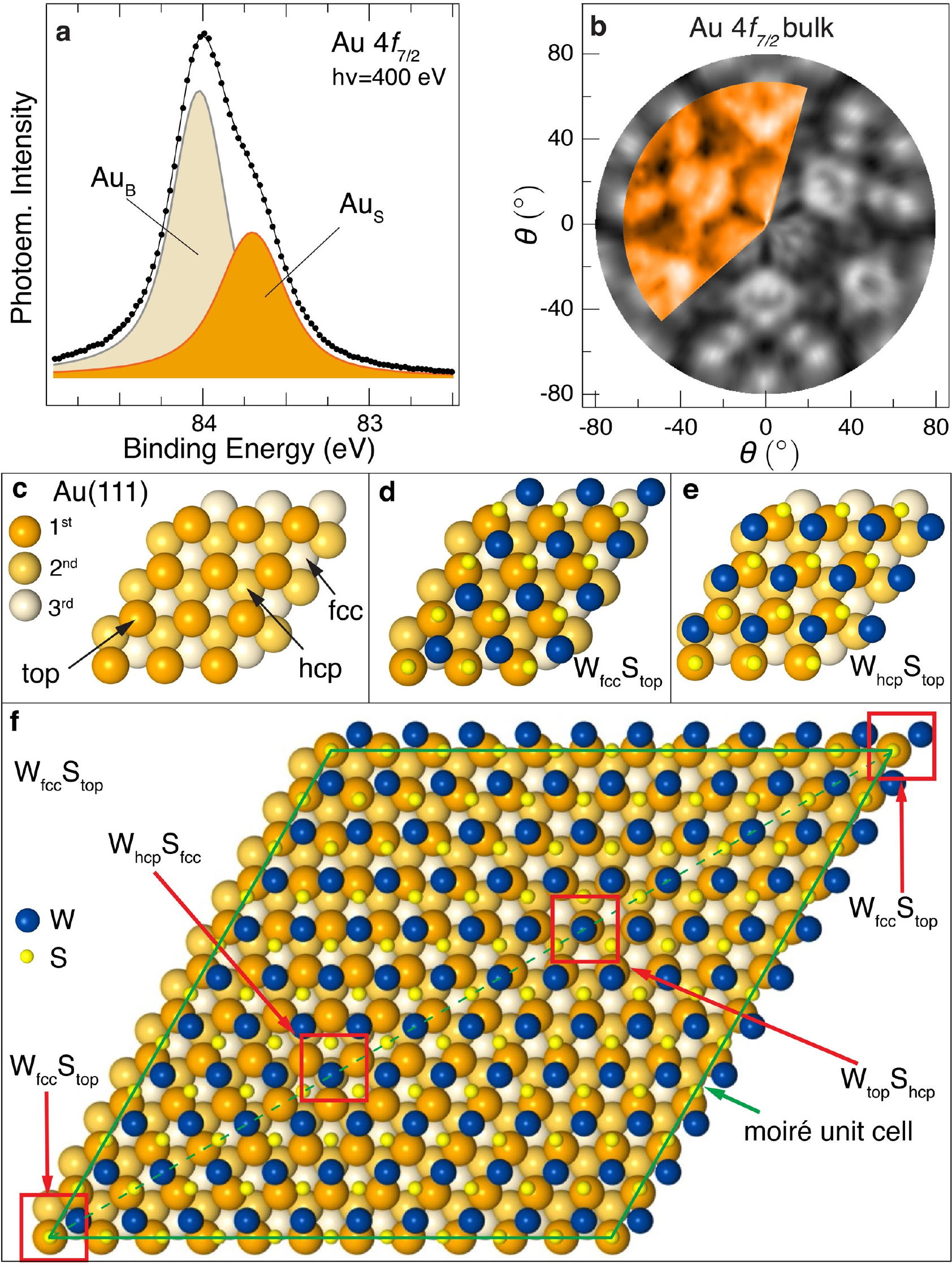}\\
\caption{(a) Au~$4f_{7/2}$ core level acquired on clean Au(111)  at $h\nu=400$~eV. Surface (Au$_\text{S}$) and bulk (Au$_\text{B}$)  components are indicated. (b)  XPD pattern acquired at 400 eV photon energy (KE$\sim$316 eV), showing the XPD pattern (color) associated to the bulk component Au$_B$ in the Au~$4f_{7/2}$ core level together with the multiple scattering simulation (grey) for the clean Au(111) sample. (b) Orientation of the Au sample derived from the comparison of the XPD experiment with the simulation. (d) and (e) Ball model of the adsorption geometry of \ce{WS2} on Au(111) at the corner of the moiré unit cell with orientation Or1 (\ce{W_{fcc}S_{top}}) and Or2 (\ce{W_{hcp}S_{top}}), respectively. (f) Moiré unit cell for the configuration Or1 (preferred) with the regions of high local symmetry for \ce{WS2} on Au(111).}
  \label{fig:5}
  \end{center}
\end{figure*}
Figure\,\ref{fig:5}a shows the typical Au~$4f_{7/2}$ spectrum acquired on clean Au(111). 
This is part of the series used to determine the XPD pattern of the bulk component Au$_B$, which is shown (colored sector) in Figure~\ref{fig:5}b together with a simulated XPD pattern (grey sector) obtained using a 4 layer slab, bulk terminated Au(111) surface, with emitters in the second and deeper layers. 
A good agreement of the three fold symmetric pattern (corresponding to the $fcc$ (111) crystal stacking) with the simulation (R-factor=0.2), returns the orientation of the Au crystal as given in Figure\,\ref{fig:5}c. Details about this R-factor analysis are given in Supplementary Material.  
Figure~\ref{fig:5}d and \ref{fig:5}e show the two possible orientations of the SL \ce{WS2} on Au(111). We call these configurations \ce{W_{fcc}S_{top}}  and \ce{W_{hcp}S_{top}}, in analogy with the case of h-BN on Ir(111)\cite{Orlando2014}. In the former configuration the sulfur atoms at the left bottom corner are adsorbed in atop position and the W atoms adsorb in the \textit{fcc} hollow sites, while in the latter W sits in \textit{hcp} hollow sites. The comparison between the XPD patterns of the Au substrate and of the \ce{WS2} overlayer unambiguously proves that the singly-oriented \ce{WS2} sample assumes the configuration \ce{W_{fcc}S_{top}} shown in Figure \ref{fig:5}d. In Figure \ref{fig:5}e we show a sketch of the entire moiré unit cell consisting of the $(10\times10)$ \ce{WS2} superstructure on the $(11\times11)$ Au unit cell for the \ce{W_{fcc}S_{top}} orientation. 
For the mixed domain growth produced with GR1, only 80\% of domains are oriented with the \ce{W_{fcc}S_{top}} configuration, while the remaining 20\% adopts the \ce{W_{hcp}S_{top}} orientation.

\begin{figure*}[!t]
\begin{center}
\includegraphics[width=.7\linewidth]{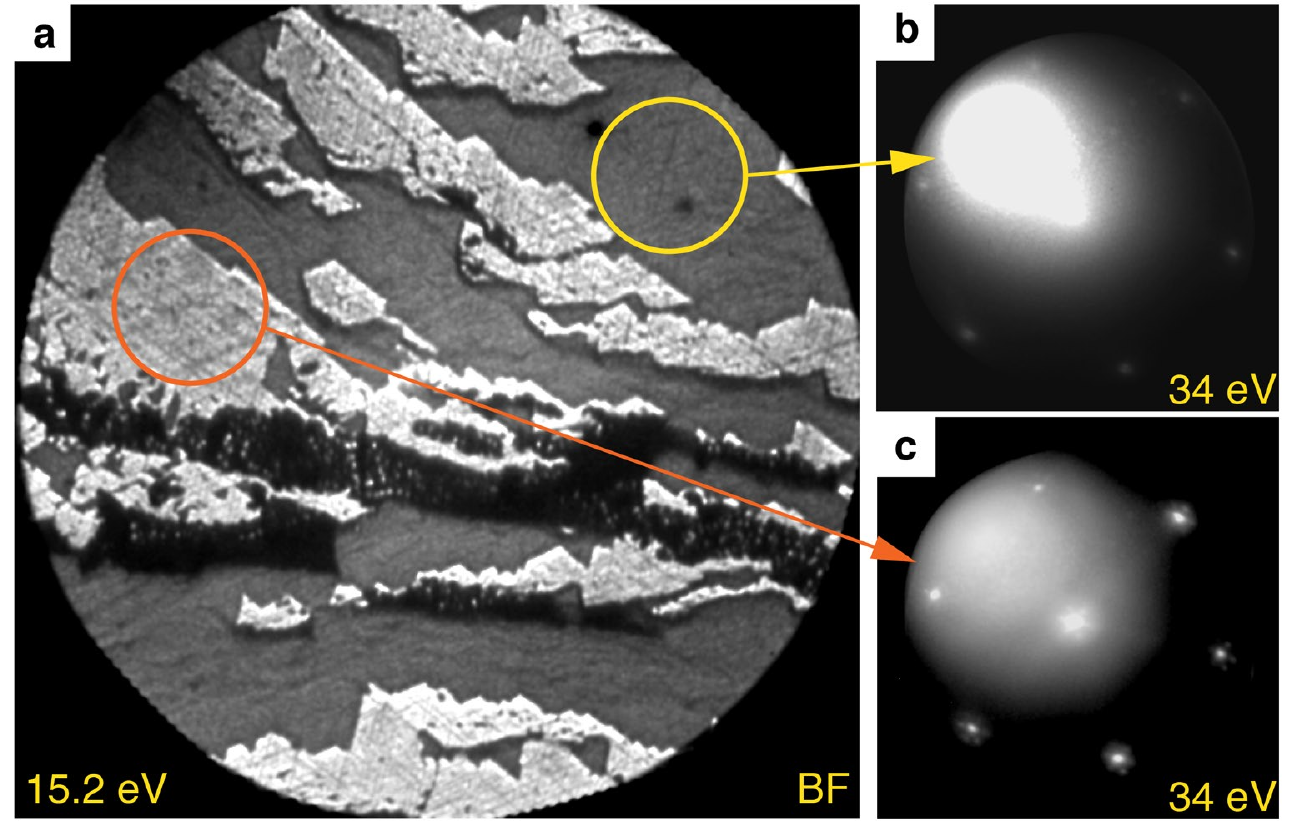}
\caption{(a) Bright-Field LEEM image (field-of-view $\sim5\mu$m, electron primary beam energy =15.2 eV) acquired on GR2, showing areas covered with SL \ce{WS2} and bare Au surface. $\mu$LEED patterns acquired on regions with (b) bare Au substrate and (c) covered with SL \ce{WS2}. }
\label{large_leem}
\end{center}
\end{figure*}
Details about the surface morphology and the size of the crystalline domains have been obtained from LEEM measurements, carried out on a  SL \ce{WS2} sample prepared ex-situ according to the GR2 procedure. In Figure \ref{large_leem}a, a typical LEEM image of the surface is shown, acquired using the specular (00) LEED beam. This imaging condition is known as bright-field (BF) mode. The surface presents a distinct contrast, distinguishing bright and dark areas. Micro-LEED patterns recorded from a 1 micron large dark area show the diffraction pattern expected for bare Au substrate (yellow circle, Figure \ref{large_leem}b) whereas the brighter areas are \ce{WS2}-covered regions (orange circle, Figure \ref{large_leem}c) with the characteristic moiré, as seen in the large area LEED pattern in Figure \ref{fig:1}e. The black areas in the center of the image can be related to surface contaminations due to transfer through air of the sample. Surface steps of the Au(111) substrate are also visible at these imaging conditions. It appears that \ce{WS2} domains have grown on individual terraces and they stretch over several microns along the steps.

\begin{figure*}[!p]
\begin{center}
\includegraphics[width=.65\linewidth]{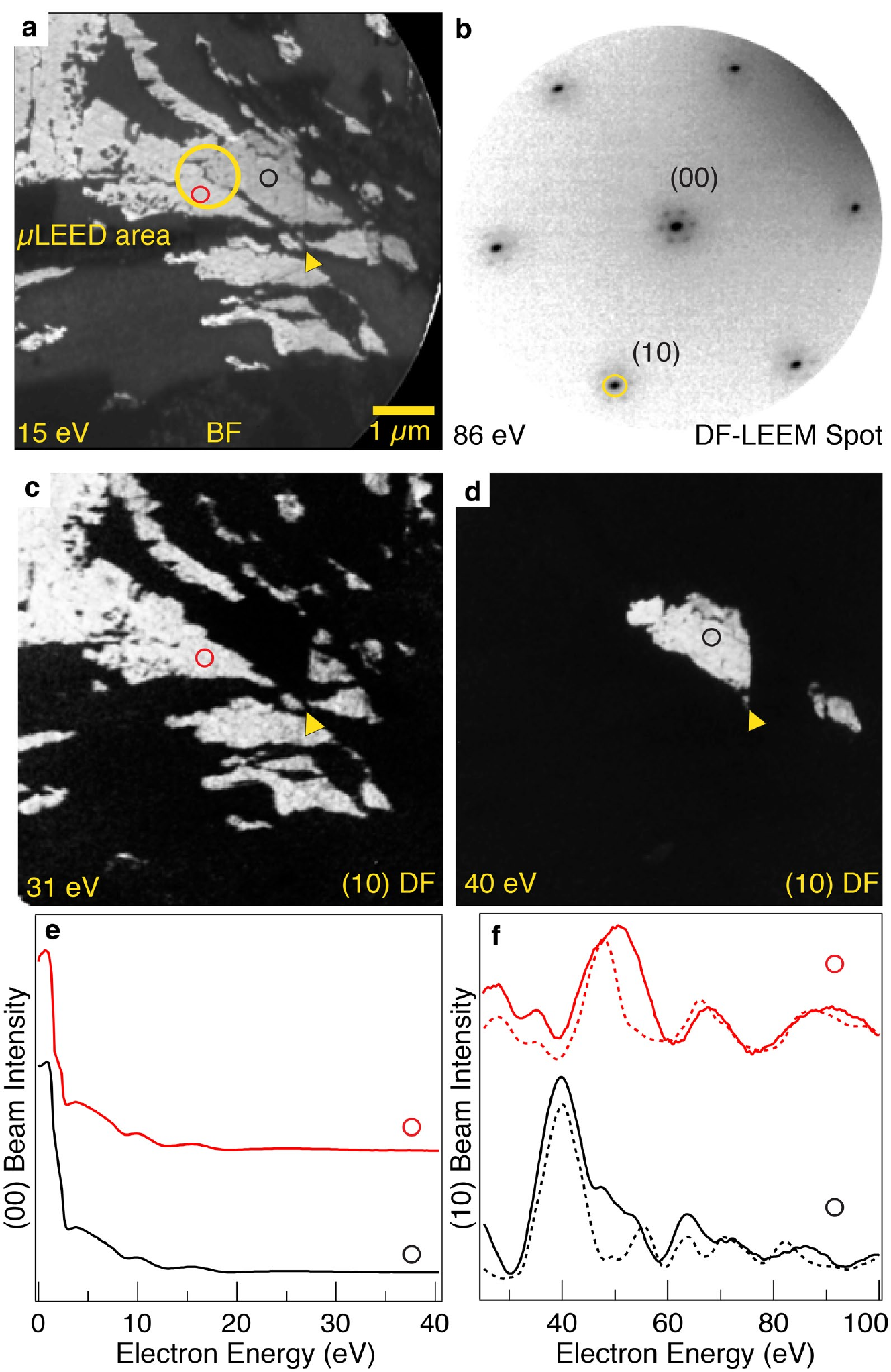}\\
\caption{(a) Bright field-LEEM reveals islands which are identified as WS$_2$ islands by their specific LEED pattern.
(b) \mbox{$\mu$LEED} pattern obtained using an  illumination aperture of 1\,$\mu$m (illuminated area marked by a yellow circle in (a)).
(c,d) Dark field LEEM images of area (a) using the (10) beam of the \ce{WS2} LEED pattern (cf. (b)), evidencing contrast changes of different \ce{WS2} domains at certain energies. The filled triangle points to a specific reference position in the image. (e) Bright field \mbox{$I(V)$ curves} from the regions marked with red and black circles. 
(f) Corresponding dark field  experimental (filled) and  simulated (dashed) \mbox{$I(V)$} curves for the two domains observed in panel (c) and (d), respectively.  The $I(V)$ curves are vertically shifted for sake of clarity. } 
  \label{leem}
  \end{center}
\end{figure*}
Figure \ref{leem}a displays a BF image acquired at larger magnification. Again, the bright areas exhibit a LEED pattern with the moiré of \ce{WS2} on Au(111) (Figure \ref{leem}b). The use of a higher order LEED beam for imaging (dark-field imaging, DF), in this case the (10) beam, should allow for contrast between rotational domains of \ce{WS2}, given the three-fold symmetry of the LEED pattern at some energies, as already shown in the LEED pattern in Figure \ref{fig:1}e. A DF image of the same surface area is presented in Figure \ref{leem}c. The majority of the \ce{WS2} domains show up bright at the chosen electron energy of 31 eV. Tuning the energy to 40 eV highlights other \ce{WS2} islands, as can be seen in Figure \ref{leem}d. We now show that the islands of different contrast correspond to opposite domain orientations. The dependence of the local LEEM intensity vs electron energy (so-called $I(V)$ LEEM \cite{Flege2014}) at the points marked in red and black in figures \ref{leem}c and \ref{leem}d can be used to determine the local structure\cite{Flege2015}. First, the $I(V)$ curves of the (00) LEED beam for the two domains are compared in Figure \ref{leem}e, revealing almost no difference. This finding directly proves that the two \ce{WS2} domains have virtually the same atomic structure \cite{Flege2014,Flege2015}. 
By contrast, the two $I(V)$ curves obtained for the (10) spot, i.e. for the DF-mode image (filled lines in Figure \ref{leem}f), are indeed  dissimilar for the two different domains. While at 40 eV one \ce{WS2} domain shows intense back reflection, at 31 eV the same domain reflects only weakly while the other does more strongly.

 In order to confirm the structural origin of these differences, we show calculated  $I(V)$ curves (dashed lines in Figure \ref{leem}f) for the two possible adsorption configurations of SL \ce{WS2}. A very good agreement between the experimental and calculated curves is found assigning the configuration \ce{W_{fcc}S_{top}} to the majority of \ce{WS2} domains (red) , while only a minor fraction of domains assumes the \ce{W_{hcp}S_{top}} (black). This interpretation is confirmed by $\mu$LEED patterns acquired in each of the two regions (see Supplementary Material). Thus, the different domain orientations can be spotted at the nanoscale because of their different $I(V)$-curve fingerprint. From an overall quantitative evaluation of the available DF images, we derive a fraction of less than 5\% of the surface being covered by the minority domains, in agreement with the results of the R-factor analysis performed for the XPD experiments. Moreover, these mirror domains are not uniformly distributed on the surface but rather were only found near surface defects, such as scratches or contaminations. This outcome sets the LEEM as an ideal technique to investigate the growth and features of TMDCs at the nanoscale, helping to understand the role of the deposition parameters and the growth mechanism from a microscopic point of view. 

\section{Conclusion}
In summary, we have reported the characterization of the growth and structural features of \ce{WS2} single layers on Au(111). 
We found that a low W deposition rate, a high partial pressure of \ce{H2S} and a high substrate temperature during the growth yield a unique crystalline orientation of SL \ce{WS2}, i.e. suppressing the presence of any detrimental mirror domains.  
Moreover, we have determined the layer stacking of \ce{WS2} on the Au(111) substrate and obtained information about the morphology and size of the domains, highlighting the differences due to the growth conditions in the structure of the final layer. 

Earlier studies about the growth of epitaxial SL \ce{WS2} did rarely explore in quantitative detail the domain morphology and orientation. However, there have been indications of a domain mixture different from 50\% for WS$_2$ on Ag(111), as circular dichroism in the excitation of a valley polarization was observed for this system  \cite{Ulstrup:2017aa}. 
We envision that the layers that we have grown and characterized in this work,  combined with methods already available to transfer \ce{WS2} from  Au to other substrates\cite{Gao2015}, can be successfully employed in valleytronics and spintronics devices and applications.

\begin{acknowledgements}
This work was supported by the Danish Council for Independent Research, Natural Sciences under the Sapere Aude program (Grant No. DFF-4002-00029) and by VILLUM FONDEN via the Centre of Excellence for Dirac Materials (Grant No. 11744). We thank Jan Lachnitt for fruitful discussion about the LEED-$I(V)$ calculations and assistance with the AQuaLEED software package. 
\end{acknowledgements}

\end{document}